\documentclass[prb,twocolumn,showpacs,preprintnumbers,amsmath,amssymb]{revtex4}

\usepackage{amsfonts}
\usepackage{latexsym}
\usepackage{graphicx}
\usepackage{dsfont}
\usepackage{epsfig}

\newcommand{\be}{\begin{equation}}
\newcommand{\ee}{\end{equation}}

\begin{document}

\title{Determination of density of states of thin high-$T_c$ films by FET type microstructures}

\author{T.M.~Mishonov}
\email[E-mail: ]{mishonov@phys.uni-sofia.bg}
\author{M.V.~Stoev}
\email[E-mail: ]{martin.stoev@gmail.com}
\affiliation{Department of Theoretical Physics, Faculty of Physics,\\
University of Sofia St.~Kliment Ohridski,\\
5 J. Bourchier Boulevard, BG-1164 Sofia, Bulgaria}

\date{\today}

\begin{abstract}
A simple electronic experiment with a field effect transistor type
microstructure is suggested. The thin superconductor layer is the
source-drain channel of the layered structure where an AC current is
applied. It is necessary to measure the second harmonic of the
source-gate voltage and third harmonic of the source-drain voltage.
The electronic measurement can give the logarithmic derivative of
the density of states which is an important parameter for fitting of
parameters of the band structures.
\end{abstract}

\pacs{74.78.-w, 71.20.-b, 74.78.Bz, 73.50.Lw}

%\preprint{Copy for friends not for distribution}

\maketitle

%%%%%%%%%%%%%%%%%%%%%%%%%%%%%%%%%%%%%%%%%%%%%%%%%%%%%%%%%%%%%%%%%%%%
\section{Introduction}
%%%%%%%%%%%%%%%%%%%%%%%%%%%%%%%%%%%%%%%%%%%%%%%%%%%%%%%%%%%%%%%%%%%%

The importance of density of states (DOS) for the physics of
high-$T_c$ cuprates was discussed in many
papers\cite{Bouvier,Bouvier and Bok, Labbe, Bouvier_Ph, Friedel,
Markiewicz, Markiewicz_JPh, Mishonov, Newns,Tsuei}. The purpose of
this work is to suggest a simple electronic method for determination
of DOS. The proposed experiment requires the preparation of
field-effect transistor (FET) type microstructure and require
standard electronic measurements. The FET controls the current
between two points but does so differently than the bipolar
transistor.  The FET relies on an electric field to control the
shape and hence the conductivity of a "channel" in a semiconductor
material. The shape of the conducting channel in a FET is altered
when a potential difference is applied to the gate terminal
(potential relative to either source or drain). It causes the
electrons flow to change it's width and thus controls the voltage
between the source and the drain. If the negative voltage applied to
the gate is high enough, it can remove all the electrons from the
gate and thus close the conductive channel in which the electrons
flow. Thus the FET is blocked.

The system, considered in this work is in hydrodynamic regime, which
means low frequency regime where the temperature of the
superconducting film adiabatically follows the dissipated Ohmic
power. All working frequencies of the lock-ins say up to 100 kHz are
actually low enough. The investigations of superconducting
bolometers show that only in MHz range it is necessary to take into
account the heat capacity of the superconducting film. As an example
there is a publication, corresponding to this topic
\cite{Mishonov:02} as well as the references therein. In this work
we propose an experiment with a FET, for which we need to measure
the second harmonic of the source-gate voltage and the third
harmonic of the source-drain voltage. Other higher harmonics will be
present in the measurements (e.g. from the leads), but in principle
they can be also used for determination of the density of states. An
analogous experimental research has been already performed for
investigation of thermal interface resistance.\cite{Chenne:03} The
suggested experiment can be done using practically the same
experimental setup, only the gate electrodes should be added to the
protected by insulator layer superconducting films.

%%%%%%%%%%%%%%%%%%%%%%%%%%%%%%%%%%%%%%%%%%%%%%%%%%%%%%%%%%%%%%%%%%%%
\section{Determination of logarithmic derivative of
density of states by electronic measurements}
%%%%%%%%%%%%%%%%%%%%%%%%%%%%%%%%%%%%%%%%%%%%%%%%%%%%%%%%%%%%%%%%%%%%

%
\begin{figure}[t]
\centering
\includegraphics[width=.9 \columnwidth]
{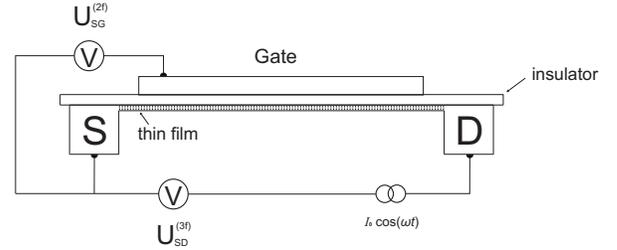} \caption{A field effect transistor(FET) is
schematically illustrated. The current $I(t)$, applied between the
source~(S) and the drain~(D) has frequency $\omega$. Running through
the transistor the electrons create voltage $U_{\mathrm{SG}}$ with
double frequency $2\omega$ between the source~(S) and the gate~(G).
The source-drain voltage $U_{\mathrm{SD}}$ is measured on the triple
frequency $3\omega$.}
\label{fig:FET}
\end{figure}
The purpose of this work is to suggest a simple electronic
experiment, determining the logarithmic derivative of the density of
states  by electronic measurements using a thin film of the material
Tl:2201. The thickness of the samples should be typical for the
investigation of high-Tc films, say 50-200 nm. Such films
demonstrate already the properties of the bulk phase. The numerical
value of this parameter
\be \nu \, '(E_F)=\frac{\mathrm{d} \nu(\epsilon)}{\mathrm{d}
\epsilon}, \ee
will ensure the absolute determination of hopping integrals.

We propose a field effect transistor (FET) from Tl:2201
Fig.~{\ref{fig:FET}} to be investigated electronically with lock-in
at second and third harmonics. Imagine a strip of Tl:2201 and
between the ends of the strip, between the source (S) and the drain
(D) is applied an AC current
\be \label{Eq:SD_current} I_\mathrm{SD}(t)=I_0\cos(\omega t). \ee

For low enough frequencies the ohmic power $P$ increases the
temperature of the film $T$ above the ambient temperature $T_0$
\be
\label{Eq:power}
P=RI_\mathrm{SD}^2=\alpha(T-T_0),
\ee
where the constant $\alpha$ determines the boundary
thermo-resistance between the Tl:2201 film and the substrate, and
$R(T)$ is the temperature dependent source-drain (SD) resistance. We
suppose that for thin film the temperature is almost homogeneous
across the thickness of the film. In such a way we obtain for the
temperature oscillations
\be \label{Eq:temperature}
(T-T_0)=\frac{RI_{SD}^2}{\alpha}=\frac{RI_0^2}{\alpha}\cos^2{(\omega
t)}. \ee
As the resistance is weakly temperature dependent
\be
\label{Eq:resistance}
R(T)=R_0 + (T-T_0) R_0\, ', \quad R_0\,
'(T_0)
    = \left.\frac{\mathrm{d} R(T)}{\mathrm{d} T}\right|_{T_0}.
\ee
A substitution here of the temperature oscillations from
Eq.~(\ref{Eq:temperature}) gives a small time variations of the
resistance
\be
\label{Eq:SD_resistance}
R(t) =
R_0\left(1+\frac{R_0'}{\alpha}I_0^2\cos^2(\omega t)\right). \ee
Now we can calculate the source-drain voltage as
\be U_{\mathrm{SD}}(t)=R(t)I_\mathrm{SD}(t). \ee
Substituting here the SD current from Eq.~(\ref{Eq:SD_current}) and
the SD resistance from Eq.~(\ref{Eq:SD_resistance}) gives for the SD
voltage
\be \label{Eq:new_U_SD}
U_\mathrm{SD}(t)=U_\mathrm{SD}^{(1f)}\cos(\omega t)
  + U_\mathrm{SD}^{(3f)}\cos(3\omega t).
\ee
The coefficient in front of the first harmonic
$U_\mathrm{SD}^{(1f)}\approx R_0 I_0$ is determined by the SD
resistance $R_0$ at low currents $I_0,$ while for the third harmonic
signal using the elementary formula $\cos^3{(\omega
t)}=(3\cos{(\omega t)}+\cos{(3\omega t)})/4$ we obtain
\be \label{Eq:3f} U_{\mathrm{SD}}^{(3f)}=
\frac{U_{\mathrm{SD}}^{(1f)}}{4\alpha}I_0^2R_0'. \ee
From this formula we can express the boundary thermo-resistance by
electronic measurements
\be \label{Eq:alpha}
\alpha=\frac{U_{\mathrm{SD}}^{(1f)}}{4U_{\mathrm{SD}}^{(3f)}}I_0^2R_0'.
\ee

The realization of the method requires fitting of $R(T)$ and
numerical differentiation at working temperature $T_0;$ the linear
regression is probably the simplest method if we need to know only
one point.

At known $\alpha$ we can express the time oscillations of the
temperature substituting in Eq.~(\ref{Eq:temperature})
\begin{equation}
 \label{Eq:temp}
T\!=\!T_0\!+\!\frac{RI_0^2}{2\alpha}\left(1+\cos(2\omega t)\right)
\!\approx \! T_0\!\left(1\!+\!
  \frac{R_{\mathrm{SD}}I_0^2}{2\alpha T_0}
  \cos (2\omega t)\! \right)\!.
\end{equation}
In this approximation terms containing $I_0^4$ are neglected and
also we consider that shift of the average temperature of the film
is small.

The variations of the temperature lead to variation of the work
function of the film according to the well-known formula from the
physics of metals
\be \label{Eq:work-function} W(T)=-\frac{\pi^2}{6e}\frac{\nu \,
'}{\nu}k_B^2T^2, \quad \nu \, '(E_F)
 =\left.\frac{\mathrm{d} \nu}{\mathrm{d} \epsilon}\right|_{E_F},
\ee
where the logarithmic derivative of the density of states
$\nu(\epsilon)$ taken for the Fermi energy $E_F$ has dimension of
inverse energy, the work function $W$ has dimension of voltage, $T$
is the temperature in Kelvins and $k_B$ is the Boltzmann constant.
For an introduction see the standard text books on statistical
physics and physics of metals.\cite{Landau, Lifshitz} Substituting
here the temperature variations from Eq.~(\ref{Eq:temp}) gives
\be \label{Eq:wtt} W=-\frac{\pi^2 k_B^2}{6e} \frac{\nu \, '}{\nu}
 T_0^2 \left[1+\frac{R_0I_0^2}{\alpha T_0}\cos(2\omega t)\right]
 +\mathcal{O}(I_0^4),
\ee
where $\mathcal{O}$-function again marks that the terms having
$I_0^4$ are negligible.

The oscillations of the temperature creates AC oscillations of the
source-gate (SG) voltage. We suppose that a lock-in with a
preamplifier, having high enough internal resistance is switched
between the source and the gate. In these conditions the second
harmonics of the work-function and of the SG voltage are equal
\be
\begin{array}{ll}
 U_\mathrm{SG}^{(2f)}= -\frac{\pi^2 k_B^2}{6e}
 \frac{\nu \,'}{\nu}T_0^2
 \frac{R_0I_0^2}{\alpha T_0},\\
  &\\ U_\mathrm{SG}(t)=U_\mathrm{SG}^{(2f)}\cos(2\omega t)
 +U_\mathrm{SG}^{(4f)}\cos(4\omega t) + \dots
\end{array}
\ee
Substituting $\alpha$ from Eq.~(\ref{Eq:alpha}) we have
\be \label{usgf} U_{\mathrm{SG}}^{(2f)}=-\frac{4\pi^2 k_B^2}{6e}
\frac{\nu{\,'}}{\nu}
\frac{U_{\mathrm{SD}}^{(3f)}}{I_0}\frac{T_0}{R_0'}. \ee
From this equation we can finally express the pursued logarithmic
derivative of the density of states
\be \label{density} \left. \frac{\mathrm{d}\ln
\nu(\epsilon)}{\mathrm{d}\epsilon}\right|_{E_F}=
\frac{\nu\,'}{\nu}=-\frac{3e}{2\pi^2k_B^2} \frac{I_0}{T_0}
\frac{U_{\mathrm{SG}}^{(2f)}}{U_{\mathrm{SD}}^{(3f)}}
\frac{d{R}}{d{T}}. \ee

In such way the logarithmic derivative of the density of states can
be determined by fully electronic measurements with a FET. This
important energy parameter can be used for absolute determination of
the hopping integrals in the generic LCAO model. The realization of
the experiment can be considered as continuation of already
published detailed theoretical and experimental investigations and
having a set of complementary researches we can reliably determine
the LCAO parameters.

\noindent \textbf{Acknowledgements} One of the authors (TM) is
thankful to J.~Bok, P.~Bernstein and J.P.~Maneval for the
stimulating discussions.


\begin{thebibliography}{99}
%
\bibitem{Friedel} J.~Friedel,
{\it J.~Phys.: Condens. Matt.} \textbf{1}, 7757 (1989);
%

%
\bibitem{Labbe} J.~Labbe and J.~Bok,
{\it Europhys. Lett.} \textbf{3}, 1225 (1987);
%

%
\bibitem{Bouvier} J.~Bouvier and J.~Bok,
{\it J.~Superconductivity} \textbf{10}, 673 (1997);
%

%
\bibitem{Bouvier and Bok} J.~Bouvier and J.~Bok,
{\it Physica} \textbf{C364-365}, 471 (2001);
%

%
\bibitem{Bouvier_Ph} J.~Bouvier and J.~Bok,
{\it Physica} \textbf{C288}, 217 (1997);
%

%
\bibitem{Markiewicz} R.~S.~Markiewicz,
{\it J.~Physics.: Condens. Matt.} \textbf{2}, 665 (1990);
%

%
\bibitem{Markiewicz_JPh} R.~S.~Markiewicz,
{\it J.~Phys. Chem. Solids} \textbf{58}, 1179-1310 (1997);
%

%
\bibitem{Newns} D.~M.~Newns, C.~C.~Tsuei and P.~C.~Pattnaik,
{\it Phys. Rev.} \textbf{52}, 13611 (1995);
%

%
\bibitem{Tsuei} C.~C.~Tsuei, C.~C.~Chi, D.~M.~Newns, P.~C.~Pattnaik
and D\"aumling, {\it Phys. Rev. Lett.} \textbf{69}, 2134 (1992);
%

%
\bibitem{Mishonov}T.~M.~Mishonov, J.~O.~Indekeu and E.~S.~Penev,
{\it J.~Phys.: Condens. Matter} \textbf{115}, 4429-4456 (2003).
%

%
\bibitem{Mishonov:02} T.M. Mishonov, N. Ch\'enne, D. Robes and J.O. Indekeu,
``Generation of 3rd and the harmonics in a thin superconducting film
by temperature and isothermal nonlinear current response'' Eur.
Phys. J. B \textbf{26}, 291-296 (2002); cond-mat/0109478.
%

%
\bibitem{Chenne:03} N. Ch\'enne, T.M. Mishonov, and J.O. Indekeu,
``Observation of a sharp lambda peak in the third harmonic voltage
response of YBaCuO film'' Eur. Phys. J. B \textbf{32}, 437-444
(2003); cond-mat/0110632.
%

%
\bibitem{Landau} L.~D.~Landau and E.~M.~Lifshitz,
{\it Statistical Physics,} Part 1, Chapter 5, (Pergamon, New York,
1977);
%

%
\bibitem{Lifshitz} I.~M.~Lifshitz, M.~Y.~Azbel and M.~I.~Kaganov
{\it Electron Theory of Metals} (Consultants Bureau, New York,
1973);
%



\end{thebibliography}
\end{document}